# New direction for gamma-rays

Eli Waxman

*The origin of energetic γ-ray bursts is still unknown. But the detection of polarization of the γ-rays provides fresh insight into the mechanism driving these powerful explosions.*

Gamma-ray bursts (GRBs) are short flashes of γ-rays, typically tens of seconds long[1]. First detected in the 1960s, GRBs are observed at a rate of roughly one per day. Although their sources are known to reside in distant galaxies, several billion light years away, what these sources are remains a mystery. But a new clue is provided by Wayne Coburn and Steven Boggs, who, on page 415 of this issue, report the detection of polarization[2] — a particular orientation of the electric-field vector — in γ-rays from a burst. This discovery may shed light on the identity of the sources of GRBs, as well as on the mechanism by which the γ-rays are produced.

The huge energy release associated with a GRB is thought to be created by the gravitational collapse of a star into a black hole or neutron star[3]. The contraction causes gravitational energy to be released. The typical energy output of a GRB corresponds to the conversion of about 1% of the Sun's mass into energy: in comparison, the energy output of an atom bomb is equivalent to the conversion to energy of approximately 1 gram of matter.

The energy released in the collapse seems to be carried away from the source in the form of a highly relativistic jet (Fig. 1), in which particles move at nearly the speed of light. As the jet reaches a radius of about 100 million kilometres from its source, part of this energy is converted to γ-rays, which become the GRB. At a later stage still, as the jet expands to a scale of 10 billion kilometres, an 'afterglow' of lower-energy radiation, at X-ray, optical and radio wavelengths, is produced.

The detection over the past few years of GRB afterglows[4] provides strong support for this scenario. But as afterglow radiation is produced at a large distance from the collapsing object, key questions remain unanswered[3]: such as, what is the nature of the collapsing object? The most popular candidate is a massive star, about ten times heavier than the Sun, whose life ends with the collapse of its core. How the gravitational energy released becomes a relativistic jet and how jet energy is converted to γ-rays are also not well understood.

From earlier observations of the γ-ray spectrum of GRBs[1], it was concluded that the most likely mechanism for γ-ray production is 'synchrotron emission' — the emission of radiation by highly energetic electrons gyrating in a strong magnetic field. But other mechanisms, such as thermal emission or energy loss by relativistic electrons in intense radiation fields, are also possibilities[3]. The radiation released through synchrotron emission is highly polarized, but the other suggested mechanisms do not naturally produce large polarization. That Coburn and Boggs detect a clear polarization in the γ-rays from a burst provides direct evidence in support of synchrotron emission as the mechanism of γ-ray production.

Their observations also reveal more about the nature of the magnetic field in which synchrotron emission occurs. In a GRB, the γ-rays are produced in different regions inside the jet (Fig. 2). These regions are unresolved by detectors close to Earth, so the source appears to be point-like. The detected polarization signal is, then, an average over the polarization of radiation produced at different points within the source. If the direction of polarization varies randomly from place to place in the jet, then the observed polarization signal is likely to average out to zero. But this 'washing out' of the polarization signal will not happen if the polarization direction is the same everywhere. For polarization produced by the synchrotron mechanism, this means that the γ-ray-producing region is suffused by an ordered magnetic field, oriented in the same direction everywhere (Fig. 2a).

The direction of polarization reported by Coburn and Boggs[2] remained constant throughout the duration of the GRB, but the γ-ray flux varied significantly. So it seems unlikely that such a strong, constant, ordered field could be generated in the region where the γ-rays are produced. Rather, this suggests that the strong field originates near the collapsing object, and is then carried by (or perhaps even drives) the jet outwards from the source: the mechanism by which gravitational energy is extracted and powers the jet is, possibly, electromagnetic.

But there could be another explanation. Strong polarization might also arise in a randomly oriented magnetic field — a structure that would be expected if the field is generated in the γ-ray production region — provided that the line-of-sight to the GRB lies close to the jet edge[5,6] (Fig. 2b). In this case, the polarization signal is not averaged out: radiation reaches the observer only from points lying to one side of the line-of-sight, closer to the jet centre; radiation from points on the other side, outside the jet cone, is 'missing'. For a highly relativistic jet, with a velocity that is 99.99% that of light, such an orientation is likely to occur by chance only if its opening angle is close to 0.01 radians (0.6 degrees); the closer the jet velocity is to the speed of light, the smaller the opening angle required. Afterglow observations suggest that jet opening angles are typically about 0.05 radians, with more powerful GRBs produced by narrower jets[7]. The burst reported by Coburn and Boggs[2] is exceptionally bright, and its jet may therefore have been very narrow. So the existence of a randomly oriented magnetic field cannot be discounted.

Future observations will determine which of these interpretations of the data is valid. Both pose challenges to models, which need to explain how a constant ordered field or a highly collimated jet might be produced.

*Eli Waxman is in the Faculty of Physics, Weizmann Institute of Science, Rehovot 76100, Israel. e-mail: waxman@wicc.weizmann.ac.il*

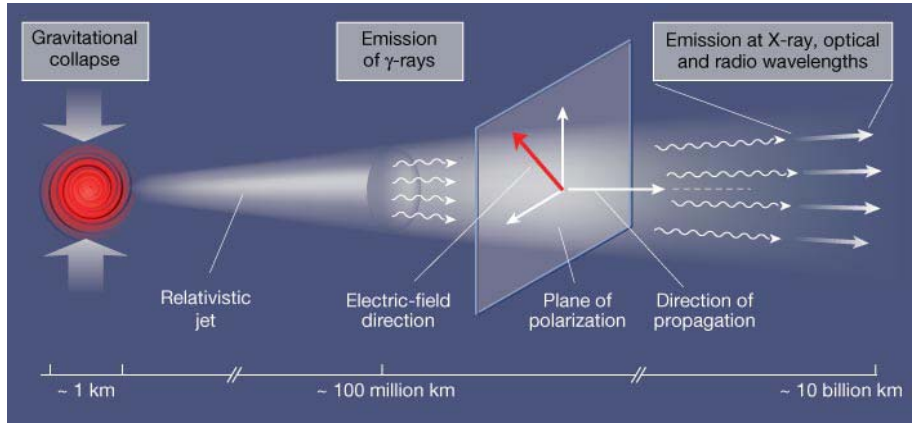

**Figure 1**
The creation of a γ-ray burst (GRB). If a dying star a few times heavier than our Sun collapsed to a diameter of just one kilometre, a large amount of gravitational energy would be released as a black hole forms. The energy is carried away by a highly relativistic jet of material, propagating at nearly the speed of light and generating, from inside the jet, a short flash of γ-rays. The electric field associated with the propagating γ-rays lies in a perpendicular plane, and may point in any direction within this 'plane of polarization'. Coburn and Boggs[2] have detected a specific orientation, or polarization, of the electric field that could be the consequence of a strong, constant and well-ordered magnetic field in the region surrounding the source of the GRB. Further from the source, the jet interacts with the surrounding medium, generating an afterglow of X-rays, optical and radio waves that lasts a few days or months.

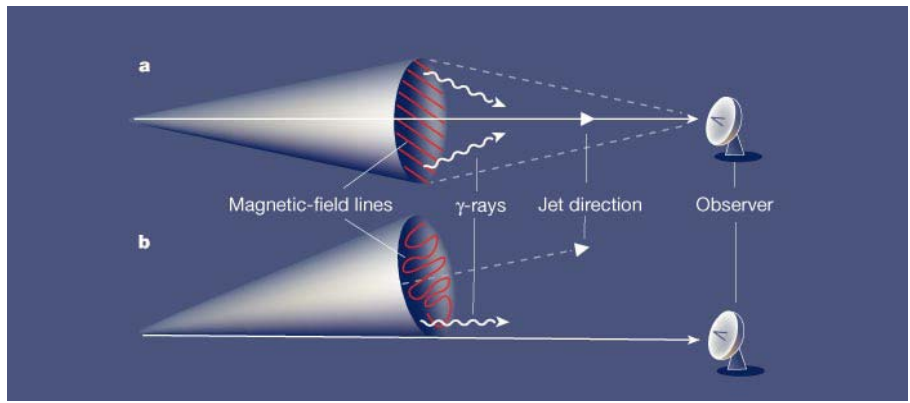

**Figure 2**
Magnetic fields and polarization. Energetic electrons gyrating in a strong magnetic field inside the jet of material ejected by a collapsing star would emit polarized γ-rays The strong degree of polarization seen by Coburn and Boggs[2] suggests that if, **a,** the observer's line-of-sight to the γ-ray burst (GRB) is close to the axis of the jet cone, the magnetic field is ordered. **b,** But if the line-of-sight to the GRB runs along the edge of the jet cone, the same degree of polarization could be seen even if the magnetic field is oriented randomly.